\documentclass[aps,floats]{revtex4}
\usepackage{amsmath,amssymb}
\usepackage{graphicx,epsfig}

\begin{document}
\bibliographystyle {plain}

\def\oppropto{\mathop{\propto}} 
\def\opsimeq{\mathop{\simeq}}
\def\opoverderline{\mathop{\overline}}
\def\operarrow{\mathop{\longrightarrow}}
\def\opsim{\mathop{\sim}}

\def\fig#1#2{\includegraphics[height=#1]{#2}}
\def\figx#1#2{\includegraphics[width=#1]{#2}}


\title{ Asymmetric scaling in large deviations  \\
for rare values bigger or smaller than the typical value    } 


\author{ C\'ecile Monthus }
 \affiliation{Institut de Physique Th\'{e}orique, 
Universit\'e Paris Saclay, CNRS, CEA,
91191 Gif-sur-Yvette, France}

\begin{abstract}

In various disordered systems or non-equilibrium dynamical models, the large deviations of some observables have been found to display different scalings for rare values bigger or smaller than the typical value. In the present paper, we revisit the simpler observables based on independent random variables, namely the empirical maximum, the empirical average, the empirical non-integer moments or other additive empirical observables, in order to describe the cases where asymmetric large deviations already occur. The unifying starting point to analyze the large deviations of these various empirical observables is given by the Sanov theorem for the large deviations of the empirical histogram : the rate function corresponds to the relative entropy with respect to the true probability distribution and it can be optimized in the presence of the appropriate constraints. Finally, the physical meaning of large deviations rate functions is discussed from the renormalization perspective.

\end{abstract}

\maketitle


\section{ Introduction }

In macroscopic systems with a large number $N$ of degrees of freedom,
it is essential to understand how physical observables fluctuate as a function of the size $N$.
When $u$ is some intensive variable, one usually distinguishes the three following levels of descriptions for its probability distribution $P_N(u)$ :

(i) in the thermodynamic limit $N \to + \infty$, the probability distribution $P_N(u) $
 becomes concentrated on the typical value $u^{typ}$ that does not depend on $N$
\begin{eqnarray}
P_N(u) \opsimeq_{N \to +\infty} \delta(u-u^{typ})
\label{largenumber}
\end{eqnarray}
This statement is the analog of the law of large numbers for the empirical average of independent random variables.
 Another well-known example 
is the typical Lyapunov exponent for product of random matrices \cite{luckbook,crisantibook}.

(ii) zooming in Eq. \ref{largenumber} will reveal the order $\frac{1}{T_N}$
of the small typical fluctuations around the typical value  $u^{typ}$.
The appropriate scale $T_N$ grows with $N$,
for instance like a power-law $N^{\chi}$ with some exponent $0<\chi<1$ or 
like a power of $\ln(N)$
\begin{eqnarray}
u \opsimeq_{N \to +\infty} u_{typ}+ \frac{v}{T_N}  
\label{vtyp}
\end{eqnarray}
 and the rescaled variable $v \equiv T_N (u-u^{typ}) $ 
is distributed with some universal limiting distribution $V(v)$.
This statement is the analog of the Central Limit Theorem with the scale $T_N = \sqrt{N}$ 
and where $V(v)$ is the Gaussian distribution
for the universality class of probability distributions whose two first moments are finite
(if they are not finite, one obtains the other universality classes involving L\'evy stable laws).
Another famous example is given by the three universality classes Gumbel-Fr\'echet-Weibull of Extreme Value Statistics \cite{Gum,Gal}, with many applications in various physics domains (see the reviews \cite{mezard,clusel,fortin} and references therein).

(iii) in the field of large deviations, one is interested instead in evaluating how rare it is for large $N$
to observe some finite value $u$ different from $u^{typ}$.
The standard theory of large deviations is based on the exponential decay \cite{oono,ellis,review_Touchette}
\begin{eqnarray}
P_N(u) \opsimeq_{N \to +\infty}  e^{ - N I (u)} 
\label{largedevusual}
\end{eqnarray}
where the rate function is positive $I (u) \geq 0$
and vanishes only for the typical value $u^{typ}$ of Eq \ref{largenumber}
\begin{eqnarray}
I(u^{typ} ) =0
\label{iratetyp}
\end{eqnarray}
While the region (ii) of universal typical fluctuations has been traditionally the main focus of studies for various physical observables,
the theory of large deviations (iii) is nowadays considered as the unifying language for the statistical physics
of equilibrium, non-equilibrium and dynamical systems (see the reviews \cite{oono,ellis,review_Touchette} and references therein).
In particular, the large deviations with respect to the large time limit of dynamical trajectories
has produced an appropriate statistical physics approach for various Markovian processes
(see the reviews  \cite{derrida-lecture,harris_Schu,searles,harris,mft,lazarescu_companion,lazarescu_generic}
and the PhD Theses \cite{fortelle_thesis,vivien_Thesis,chetrite_Thesis,wynants_Thesis} 
 and the HDR Thesis \cite{chetrite_HDR}).

However the recent huge activity on large deviations in the field of random matrices
has shown that the maximal eigenvalue \cite{dean_maj,EVSbouchaud,maj_verg,satya,gregory,cauchy}
and many other observables involving the eigenvalues 
\cite{oriol,celine,celine_ent,pca,christophe,marino,marino2,aurelien16,shortcut,aurelien17,lacroix}
display asymmetric scaling in large deviations : 
 the probability $P_N(u) $ to observe bigger values than typical 
$u> u^{typ}$ and smaller values than typical $ u< u^{typ}$ are governed by two different scalings $D_N^{\pm}$ (for instance two different power-laws $D_N^{\pm} =N^{\theta_{\pm}} $) and two rate functions $I_{\pm}(u)$
\begin{eqnarray}
P_N(u) \opsimeq_{N \to +\infty}  e^{ -D_N^+ I_+ (u)} \ \ {\rm for  \ } u\geq u^{typ}
\nonumber \\
P_N(u) \opsimeq_{N \to +\infty}  e^{ -D_N^- I_- (u)} \ \ {\rm for  \ } u\leq u^{typ}
\label{largedevpm}
\end{eqnarray}
instead of the standard form of Eq. \ref{largedevusual}.
For the maximal eigenvalue \cite{dean_maj,EVSbouchaud,maj_verg,satya,gregory,cauchy},
 the physical interpretation of this asymmetry is that 
to push the maximal eigenvalue inside the Wigner sea, one needs to reorganize all the other eigenvalues,
whereas to pull the maximal eigenvalue outside the Wigner sea, one may leave the other eigenvalues
unchanged.
Via mapping between models belonging to the Kardar-Parisi-Zhang universality class
(see the list in the review \cite{gregory} and references therein),
these asymmetric large deviations properties for the biggest eigenvalue
of some random matrices ensembles
can be rephrased in many other frameworks, in particular :

(a) for the Asymmetric Exclusion process, which is one of the most studied models in the field of the non-equilibrium dynamics of interacting particles (see the reviews \cite{derrida-lecture,lazarescu_companion,lazarescu_generic}
 and references therein), the interpretation of the asymmetric large deviations is that to slow down the traffic, it is sufficient to slow down a single particle, 
whereas to speed up the traffic, one needs to speed up all particles \cite{derrida_leb}. 

(b) for the Directed Polymer in random medium in dimension $d=2$, which is one
 of the simplest disordered model 
displaying a low temperature glassy frozen phase (see the review \cite{Hal_Zha} and references therein), 
the interpretation is that an anomalously good ground state energy requires only $L$ anomalously good on-site energies along the polymer,
while an 'anomalously bad' ground state energy requires $L^d$ bad on-site energies in the sample. 

These examples and their very clear physical meanings
show that asymmetric large deviations of Eq. \ref{largedevpm}
are likely to occur in many other problems in the fields of non-equilibrium dynamics or disordered systems,
while they are not considered in the standard theory of large deviations \cite{oono,ellis,review_Touchette}
based on Eq. \ref{largedevusual}.
As a consequence, 
it seems useful to revisit simpler observables based on independent random variables
where asymmetric large deviations have been found to occur,
 in particular for the empirical maximum \cite{maxmath,vivo},
for the empirical average \cite{nagaev,evans2008,nina,godreche},
and in joint linear statistics \cite{evans2014}.
Since these problems have been already studied in details in these references by exact methods,
 the goal of the present paper is 
to give a unifying perspective based on the large deviation properties of the empirical histogram 
in the presence of constraints corresponding to the observables under study.
This point of view also allows to make the link with the studies of large deviations in the field of random matrices
\cite{dean_maj,EVSbouchaud,maj_verg,satya,gregory,cauchy,oriol,celine,celine_ent,pca,christophe,
marino,marino2,aurelien16,shortcut,aurelien17,lacroix}
where the Coulomb gas technique is based on the large deviations of the empirical histogram of eigenvalues,
in the presence of constraints corresponding to the observables under study.
The main difference is that the large deviations of the empirical histogram are
governed by the Coulomb interaction energy in the case of random matrices eigenvalues,
while it is governed by the relative entropy of the Sanov theorem for the case of independent variables  \cite{oono,ellis,review_Touchette}.
This large deviation framework also makes the link with 
the Gibbs theory of ensembles in equilibrium statistical physics 
 \cite{oono,ellis,review_Touchette} and thus allows to understand 
why it is natural to expect the possibility of phase transitions
in large deviation rate functions (see the recent review \cite{largedevsing} and references therein).

The paper is organized as follows.
In section \ref{sec_nota}, we recall how the empirical histogram of independent random variables
allows to reconstruct interesting observables like the empirical maximum, the empirical average, or
other additive empirical observables, with the consequences for typical values.
In section \ref{sec_largedevhisto}, the large deviations of the empirical histogram is presented as
the unifying starting point to analyze the large deviations of empirical observables.
In section \ref{sec_max}, the asymmetry in the large deviations of the empirical maximum 
is analyzed on various scales.
In section \ref{sec_average}, the asymmetry in the large deviations of the empirical average
is described for the case of stretched exponential decay or power-law decay of the initial distribution.
In section \ref{sec_momentsq}, the generalization for the large deviations 
of arbitrary non-integer empirical moments is discussed.
In section \ref{sec_rg}, these large deviations properties
are analyzed from the renormalization perspective.
Our conclusions are summarized in section \ref{sec_conclusion}.


\section{ Empirical observables for independent random variables }

\label{sec_nota}

\subsection{ Notations }

Our main goal is to analyze the asymmetry in large deviation properties
that may occur for simple observables involving $N$ independent random variables $x_i $
 drawn with some probability distribution $\pi(x)$.
To simplify the discussion, we will
 focus on the cases of positive variables $0 \leq x < +\infty$,
where the decay of the probability distribution $\pi(x)$ for large $x \to +\infty $ 
is :

(i) either an exponential decay with some exponent $\alpha >0 $
\begin{eqnarray}
\pi^{exp}_{\alpha,\nu} (x) \opsimeq_{x \to +\infty} K x^{\nu-1} e^{-x^{\alpha} } 
\label{expalpha}
\end{eqnarray}
 with possibly some power-law prefactor if $\nu \ne 1$,
while $K$ is a constant amplitude.

ii) or a power-law decay with some exponent $\mu>2$
(in order to ensure the existence of the two first moments $\overline{x}  $ and $\overline{x^2}  $)
\begin{eqnarray}
\pi^{power}_{\mu}(x) \opsimeq_{x \to +\infty} \frac{K  }{x^{1+\mu} }
\label{powermu}
\end{eqnarray}

But of course these assumptions are not restrictive, 
and if one is interested into other cases,
one can easily adapt the methods described below by considering the various possible tail behaviors for $x \to - \infty$.

\subsection{ Empirical histogram }

If one is not interested in the order of appearance of the variables $[x_i]_{1 \leq i \leq N}$ (otherwise see the pedagogical introduction \cite{largedevdisorder} and references therein), all the information is contained in the empirical histogram
\begin{eqnarray}
p_N(x) \equiv \frac{1}{N} \sum_{i=1}^N \delta(x-x_i)
\label{empihisto}
\end{eqnarray}
Its typical value is of course the 'true' probability distribution $\pi(x)$
\begin{eqnarray}
p_N^{typ}(x) = \pi(x)
\label{empihistotyp}
\end{eqnarray}
The large deviations around this typical value will be discussed in section \ref{sec_largedevhisto}.
Let us first recall how the empirical histogram allows to reconstruct the usual empirical observables of interest.

\subsection{ Empirical maximum  }

The information on the empirical maximum
\begin{eqnarray}
x^{max}_N \equiv \max \limits_{1 \leq i \leq N}  \left( x_i  \right)
\label{empimax}
\end{eqnarray}
is contained in the empirical histogram of Eq. \ref{empihisto} as follows :
the empirical maximum $x^{max}_N$ corresponds to the value $x$ 
where the empirical number of variables bigger than $x$
\begin{eqnarray}
{\cal N}_N(x ) \equiv \sum_{i=1}^N \theta(x_i-x) = N \int_x^{+\infty} dy p_N(y)
\label{empiricalbigger}
\end{eqnarray}
jumps from the value $0$ for $x$ bigger than $x^{max}_N $
 towards the value $1$ for $x$ sligthly smaller then $x^{max}_N  $
\begin{eqnarray}
0={\cal N}_L(x^{max}_N+0 ) =N \int_{x^{max}_N+0}^{+\infty} dy p_N(y)
\nonumber \\
1={\cal N}_L(x^{max}_N-0 ) =N \int_{x^{max}_N-0}^{+\infty} dy p_N(y)
\label{maxhisto}
\end{eqnarray}

The typical value of the empirical histogram of Eq. \ref{empihistotyp}
yields that
the typical value $M_N$ of the empirical maximum $x_N^{max} $ of Eq. \ref{empimax}
\begin{eqnarray}
M_N \equiv \left( x_N^{max} \right)^{typ}
\label{mn}
\end{eqnarray}
is given by the typical position of the jump of Eq. \ref{maxhisto}
\begin{eqnarray}
1=N \int_{M_N}^{+\infty} dy p_N^{typ}(y) = N \int_{M_N}^{+\infty} dy \pi(y) \equiv N C(M_N)
\label{maxtyp}
\end{eqnarray}
where we have introduced the
complementary cumulative distribution function that measures the integrated tail above 
the threshold $x$
\begin{eqnarray}
C(x) \equiv \int_{x }^{+\infty}  dx' \pi(x')
\label{cumulative}
\end{eqnarray}

The large deviations properties of the ratio
\begin{eqnarray}
r_N \equiv \frac{x_N^{max} }{M_N} = \frac{1}{M_N}\left( \max \limits_{1 \leq i \leq N}  \left( x_i  \right) \right)
\label{maxratio}
\end{eqnarray}
of typical value unity
\begin{eqnarray}
r_N^{typ}=1
\label{rntyp}
\end{eqnarray}
will be discussed in section \ref{sec_max}.
To be self-contained, let us now recall the behavior of the typical values $M_N$ as a function of $N$
for the two types of decay under study here.

\subsubsection{ Typical value $M_N$ of the maximum for the exponential decay }

For the asymptotic behavior of Eq. \ref{expalpha},
the asymptotic behavior of its primitive of Eq. \ref{cumulative} reads
\begin{eqnarray}
C(x)   \opsimeq_{x \to +\infty} \frac{K}{\alpha} x^{\nu-\alpha} e^{-x^{\alpha} } 
\label{expcumul}
\end{eqnarray}
Then Eq. \ref{maxtyp} determining the typical value $M_N$ of the empirical maximum
becomes for large $N$
\begin{eqnarray}
\frac{1}{N}  = C(M_N) \opsimeq_{N \to +\infty}  \frac{K}{\alpha} M_N^{\nu-\alpha} e^{-M_N^{\alpha} } 
\label{defxnstareta}
\end{eqnarray}
The inversion yields at leading order the well-known logarithmic behavior \cite{Gum,Gal}
\begin{eqnarray}
M_N \opsimeq_{N \to +\infty} \left[ \ln N + \frac{\nu-\alpha}{\alpha} \ln (\ln N) - \ln \left( \frac{\alpha}{K} \right) \right]^{\frac{1}{\alpha}} 
\label{mntypexp}
\end{eqnarray}

\subsubsection{ Typical value $M_N$ of the maximum  the power-law decay }

For the power-law decay of Eq. \ref{powermu},
the asymptotic behavior of its primitive of Eq. \ref{cumulative}
\begin{eqnarray}
C(x) = \int_{x}^{+\infty} dx' \pi(x')= \frac{K}{\mu x^{\mu} } 
\label{cumulpower}
\end{eqnarray}
yields that the solution of Eq. \ref{maxtyp} follows the well-known power-law \cite{Gum,Gal}
\begin{eqnarray}
M_N \ =\left( \frac{ KN}{\mu} \right)^{\frac{1}{\mu}} 
\label{mntyppower}
\end{eqnarray}

\subsection{ Empirical additive observables  }

The empirical histogram of Eq. \ref{empihisto} allows to reconstruct 
any additive observable $G_N$ involving some function $g(x)$
\begin{eqnarray}
G_N \equiv \frac{1}{N} \sum_{i=1}^N g(x_i) = \int_0^{+\infty} dx g(x) p_N(x)
\label{gadditive}
\end{eqnarray}
Th most studied observable in the whole history of probability is of course
 the empirical average
\begin{eqnarray}
a_N \equiv \frac{1}{N} \sum_{i=1}^N x_i = \int_0^{+\infty} dx x p_N(x)
\label{empiav}
\end{eqnarray}
The empirical moments of arbitrary non-integer order $q$
\begin{eqnarray}
a^{(q)}_N \equiv \frac{1}{N} \sum_{i=1}^N x_i^q = \int_0^{+\infty} dx x^q p_N(x)
\label{empimomentsq}
\end{eqnarray}
have been also considered \cite{EVSbouchaud,clusel} in order to interpolate between
the case $q=1$ of the empirical average of Eq. \ref{empiav} and the empirical maximum of Eq. \ref{empimax}
that should dominate the empirical moment of Eq \ref{empimomentsq} for large $q \to +\infty$.
Another examples include the exponential case $g(x)=e^{tx}$ considered in Ref \cite{EVSbenarous}
or the logarithmic case $g(x)=\ln (x)$.

The typical value of the empirical histogram of Eq. \ref{empihistotyp}
yields that the typical values of additive observables of Eq. \ref{gadditive}
are simply
\begin{eqnarray}
G_N^{typ}  = \int_0^{+\infty} dx g(x) p^{typ}_N(x) = \int_0^{+\infty} dx g(x)\pi(x)
\label{gadditivetyp}
\end{eqnarray}
In particular the typical value of the empirical average of Eq. \ref{empiav}
corresponds to the first moment $ \overline{x}$ of $\pi(x)$
\begin{eqnarray}
a_N^{typ} =\int_0^{+\infty} dx x p^{typ}_N(x) = \int_0^{+\infty} dx x \pi(x) = \overline{x}
\label{empiavtyp}
\end{eqnarray}
The possibility of asymmetric 
large deviations properties around this typical value will be discussed in section \ref{sec_average}.


\section{ Analysis based on the large deviations of the empirical histogram }

\label{sec_largedevhisto}

\subsection{ Reminder on the Sanov theorem involving the relative entropy  }

The large deviations of the empirical histogram $p_N(x)$ of Eq. \ref{empihisto}
around its typical value $p_N^{typ}(x)=\pi(x) $ of Eq \ref{empihistotyp}
are described by the Sanov theorem (see the reviews \cite{oono,ellis,review_Touchette}
and the pedagogical introduction \cite{largedevdisorder})
\begin{eqnarray}
{\cal P}_N [ p_N(.)  ]   \opsimeq_{N \to +\infty} \delta \left( 1-  \int dx p_N(x)    \right)
e^{ \displaystyle - N S^{rel} ( p_N(.)  \vert \pi(.))  }
\label{sanov}
\end{eqnarray}
The delta function appears in order to impose the normalization constraint of the empirical histogram $  \int dx p_N(x) =1 $.
The exponentially small term in $N$ involves the relative entropy of the empirical histogram  $p_N(x)$
with respect to the true probability distribution $\pi(x)$ 
\begin{eqnarray}
S^{rel} ( p_N(.)  \vert \pi(.))   \equiv \int dx  p_N(x)  \ln \left( \frac{ p_N(x)  }{ \pi(x)   } \right)
\label{s1relative}
\end{eqnarray}

\subsection{ Exact generating function of the empirical histogram for finite $N$ }

Among the various derivations of Eq. \ref{sanov}, one is based on 
 the exact generating function ${\cal Z}_N [ \kappa(.)  ] $ of the empirical histogram for any finite $N$
\begin{eqnarray}
 {\cal Z}_N [ \kappa(.)  ]  && 
\equiv  \int {\cal D}{ p_N(.) }  {\cal P}_N [ p_N(.)  ]    \ e^{ \displaystyle  N \int dx \kappa(x) p_N(x) }   
 = \int dx_1 \pi(x_1) ... \int dx_N \pi(x_N)
 \  e^{ \displaystyle   \sum_{i=1}^N \kappa(x_i) }   
 \nonumber \\ &&
 = \prod_{i=1}^N \left(  \int dx_i \pi(x_i)  e^{\kappa(x_i) } \right)
= \left(  \int dx \pi(x)  e^{\kappa(x) }     \right)^N \equiv e^{ N \Phi [ \kappa(.) ]  }
\label{geneq1}
\end{eqnarray}
where the scaled cumulant generating function 
\begin{eqnarray}
 \Phi [ \kappa(.)  ]  && = \ln \left(  \int dx \pi(x)  e^{\kappa(x) }     \right)
\label{phi1}
\end{eqnarray}
is related to the relative entropy of Eq. \ref{s1relative}
via the appropriate Legendre transform (see \cite{largedevdisorder} for more details on the Legendre transforms in the two directions).

\subsection{ Constraint to reproduce the cumulative distribution of the maximum $x_N^{max}$ }

The probability distribution ${\cal X}_N(x^{max}_N) $ of the empirical maximum $x^{max}_N $ of Eq. \ref{empimax} is well known to be \cite{Gum,Gal}
\begin{eqnarray}
 {\cal X}_N(x^{max}_N)  = N  \pi(x^{max}_N ) \left[ 1- C(x^{max}_N)\right]^{N-1}
\label{pxmax}
\end{eqnarray}
in terms of the cumulative function $C(x)$ introduced in Eq. \ref{cumulative}.

Here it is instructive to mention how it can be reproduced from
the large deviations of the empirical histogram of Eq. \ref{sanov} : 
the cumulative probability of the maximum $x_N^{max}$ amounts to 
replace the normalisation constraint $ \int dx p_N(x) =1  $ by the two constraints
\begin{eqnarray}
\int_{x_N^{max}}^{+\infty} dx p_N(x) && =0
\nonumber \\
\int_0^{x_N^{max}} dx p_N(x) && =1
\label{constraintsmax}
\end{eqnarray}
leading to
\begin{eqnarray}
 \int_0^{x_N^{max}} dx {\cal X}_N(x)  && = \int {\cal D} p_N(.)    \delta \left( 1-  \int_0^{x_N^{max}} dx p_N(x)    \right)
\delta \left( \int_{x_N^{max}}^{+\infty} dx p_N(x) \right)
e^{ \displaystyle - N S^{rel} ( p_N(.)  \vert \pi(.))  }
\label{cumulsanov}
\end{eqnarray}
Introducing the Lagrange multiplier $\omega$, one needs to optimize the functional
\begin{eqnarray}
{\cal L} (p_N(.))  && =  -  S^{rel} ( p_N(.)  \vert \pi(.))   +\omega  \left( 1-  \int_0^{x_N^{max}} dx p_N(x)    \right)
\nonumber \\
&& = - \int dx  p_N(x)  \ln \left( \frac{ p_N(x)  }{ \pi(x)   } \right)+\omega  \left( 1-  \int_0^{x_N^{max}} dx p_N(x)    \right)
\label{lmax}
\end{eqnarray}
over the empirical histogram $ p_N(.)$
\begin{eqnarray}
0 = \frac{\partial {\cal L} (p_N(.)) }{\partial p_N(x) }  
= -   \ln \left( \frac{ p_N(x)  }{ \pi(x)   } \right)-1 -\omega  
\label{derilmax}
\end{eqnarray}
The optimal solution is thus simply proportional to the true distribution $\pi(x) $ on $[0,x_N^{max}]$
(while it vanishes for $x>x_N^{max}$)
\begin{eqnarray}
p^*_N(x) = \pi(x)  e^{-1 -\omega  } \theta(0 \leq x \leq x_N^{max} )
\label{optimax}
\end{eqnarray}
where the normalization constraint determines the Lagrange multiplier $\omega$
\begin{eqnarray}
1 = \int_0^{x_N^{max}} dx p^*_N(x) =e^{-1 -\omega  }   \int_0^{x_N^{max}} dx \pi(x)  = e^{-1 -\omega  } \left[ 1- C(x^{max}_N)\right]
\label{omegamax}
\end{eqnarray}
Plugging the corresponding optimal value of the functional of Eq. \ref{lmax}
\begin{eqnarray}
{\cal L} (p^*_N(.))  
&& = - \int dx  p^*_N(x)  \ln \left( \frac{ p^*_N(x)  }{ \pi(x)   } \right)
= 1+\omega = \ln \left[ 1- C(x^{max}_N)\right]
\label{lmaxopt}
\end{eqnarray}
into Eq. \ref{cumulsanov}
\begin{eqnarray}
 \int_0^{x_N^{max}} dx {\cal X}_N(x)  && \opsimeq_{N \to +\infty} e^{N  {\cal L} (p^*_N(.))  } =  e^{N \ln \left[ 1- C(x^{max}_N)\right]  }
\label{cumulsanovres}
\end{eqnarray}
thus allows to recover the exact cumulative distribution of Eq. \ref{pxmax}.
In this derivation, Eq. \ref{cumulsanov} thus corresponds to the entropic cost for 
the emptiness of the region $[x_N^{max},+\infty[$.
Section \ref{sec_max} will be devoted to the asymmetric large deviations properties of this distribution.

\subsection{ Standard large deviations for additive empirical observables }

\label{sec_standardLargedev}

The probability distribution ${\cal G}_N(G_N) $ of the additive observable of Eq. \ref{gadditive} 
\begin{eqnarray}
G_N \equiv \frac{1}{N} \sum_{i=1}^N g(x_i) = \int_0^{+\infty} dx g(x) p_N(x)
\label{gadditivebis}
\end{eqnarray}
can be directly characterized by its exact generating function for finite $N$ 
by applying Eq \ref{geneq1} to the case $\kappa(x)= k g(x) $
\begin{eqnarray}
 Z_N (k)  && 
\equiv  \int dG {\cal G}_N(G) e^{ \displaystyle  N k G }   
= \int {\cal D}{ p_N(.) }  {\cal P}_N [ p_N(.)  ]    \ e^{ \displaystyle  N k \int dx g(x) p_N(x) }   
= \left(  \int dx \pi(x)  e^{k g(x) }     \right)^N \equiv e^{ N \phi(k)  }
\label{geneg}
\end{eqnarray}
with the scaled cumulant generating function $\phi (k)$ 
\begin{eqnarray}
 \phi(k)  && = \ln \left(  \int dx \pi(x)  e^{ k g(x) }     \right)
\label{phig}
\end{eqnarray}
The alternative evaluation of Eq. \ref{geneg} 
based on the standard large deviation form for the probability 
\begin{eqnarray}
{\cal G}_N(G )  && \opsimeq_{N \to \infty} e^{ -N I (G)} 
\label{largedevusualg}
\end{eqnarray}
yields 
\begin{eqnarray}
e^{ N \phi(k) } \equiv \int dG {\cal G}_N(G) e^{ \displaystyle  N k G }   
\opsimeq_{N \to +\infty}\int dG   e^{ N \left( k G -I(G) \right)} 
\label{expnphik}
\end{eqnarray}
via the saddle-point method for the integral over $G$ that
$\phi(k)$ is the Legendre transform of the rate function $I(G)$ 
\begin{eqnarray}
\phi(k)  && = k G -I(G) 
\nonumber \\
0 && = k - I'(G)
\label{legendre}
\end{eqnarray}
with the reciprocal Legendre transform
\begin{eqnarray}
I(G)   && = k G  - \phi(k) 
\nonumber \\
0 && = G   - \phi'(k) 
\label{legendrereci}
\end{eqnarray}

Another way to understand the physical meaning of these Legendre transforms
consists in evaluating the probability $  {\cal G}_N(G) $
via the addition of the sum constraint in
the large deviations of the empirical histogram of Eq. \ref{sanov} : 
\begin{eqnarray}
 {\cal G}_N(G)  && = \int {\cal D} p_N(.)    \delta \left( 1-  \int_0^{+\infty} dx p_N(x)    \right)
\delta \left( G- \int_{0}^{+\infty} dx g(x) p_N(x) \right)
e^{ \displaystyle - N S^{rel} ( p_N(.)  \vert \pi(.))  }
\label{gsanov}
\end{eqnarray}
Introducing the two Lagrange multiplier $\omega$ and $k$, one needs to optimize the functional
\begin{eqnarray}
{\cal L} (p_N(.))  && =  -  S^{rel} ( p_N(.)  \vert \pi(.))  
 +\omega  \left( 1-  \int_0^{+\infty} dx p_N(x)    \right) 
-k  \left( G-  \int_0^{+\infty} dx g(x) p_N(x)    \right) 
\nonumber \\
&& = - \int dx  p_N(x)  \ln \left( \frac{ p_N(x)  }{ \pi(x)   } \right)
+\omega  \left( 1-  \int_0^{+\infty} dx p_N(x)    \right) 
-k  \left( G-  \int_0^{+\infty} dx g(x) p_N(x)    \right) 
\label{lg}
\end{eqnarray}
over the empirical histogram $ p_N(.)$
\begin{eqnarray}
0 = \frac{\partial {\cal L} (p_N(.)) }{\partial p_N(x) }  
= -   \ln \left( \frac{ p_N(x)  }{ \pi(x)   } \right)-1 -\omega  +k g(x)
\label{derilg}
\end{eqnarray}
The optimal solution reads
\begin{eqnarray}
p^*_N(x) =e^{-1 -\omega +k g(x) }  \pi(x)  
\label{optig}
\end{eqnarray}
where the the Lagrange multipliers are fixed by the two constraints
\begin{eqnarray}
1 && = \int_0^{+\infty} dx p^*_N(x) = e^{-1 -\omega } \int_0^{+\infty} dx e^{k g(x) }  \pi(x)  
\nonumber \\
G && = \int_0^{+\infty} dx g(x) p^*_N(x) = e^{-1 -\omega } \int_0^{+\infty} dx g(x) e^{k g(x) }  \pi(x)  
\label{omegagg}
\end{eqnarray}
i.e. in terms of the function $\phi(k) $ introduced in Eq. \ref{phig}
\begin{eqnarray}
1+\omega && =\phi(k)  
\nonumber \\
G && = \phi'(k)
\label{multiplierphig}
\end{eqnarray}
The corresponding optimal value of the functional of Eq. \ref{lg}
using \ref{multiplierphig} thus involves the Legendre transform $I(G)$ of $\phi(k)$
(Eqs \ref{legendre} and \ref{legendrereci})
\begin{eqnarray}
{\cal L} (p^*_N(.))  
&& = - \int dx  p^*_N(x)  \ln \left( \frac{ p^*_N(x)  }{ \pi(x)   } \right)
= 1+\omega - k G = \phi(k) - k G = - I(G)
\label{lgopt}
\end{eqnarray}
as it should to recover via Eq. \ref{gsanov}
\begin{eqnarray}
 {\cal G}_N(G)  && \opsimeq_{N \to +\infty} 
e^{ \displaystyle  N {\cal L} (p^*_N(.))    } = e^{ \displaystyle - N I(G)    } 
\label{gsanovres}
\end{eqnarray}

Here the analogy with 
the Gibbs theory of ensembles in equilibrium statistical physics is obvious :
the effective distribution of Eq \ref{optig} for an individual random variable $x$
is the analog of the Boltzmann distribution in the canonical ensemble,
where the Lagrange multiplier $k$ is conjugated to the quantity $g(x)$,
whose average $G$ over the $N$ variables is fixed.

Of course these computations make sense only if the integrals of Eq. \ref{omegagg} converge :
depending on the function $g(x)$ defining the empirical observable $G_N$ under study,
these integrals may diverge in some region of the Lagrange multiplier $k$.
The consequences for the non-standard large deviations properties
will be discussed for the case of the empirical average in section \ref{sec_average}
and for the empirical moments in section \ref{sec_momentsq}.

\subsection{ Large deviations for joint additive empirical observables }

If one is interested in the large deviations of the joint probability of two additive empirical observables,
one needs to add another constraint in Eq. \ref{gsanov},
as described in detail in Ref \cite{evans2014} for the joint probability
of the empirical average and of an empirical moment of order $q$.
More generally, one can add as many constraints as needed for the problem one is interested in.

\section{ Asymmetry in the Large deviations of the empirical maximum }

\label{sec_max}

\subsection{ Probability distribution of the ratio $r_N =  \frac{x_N^{max} }{M_N}$ }

Via the change of variables $r =  \frac{x_N^{max} }{M_N}$ of Eq. \ref{maxratio}, 
the probability distribution ${\cal X}_N(x^{max}_N)  $ of Eq. \ref{pxmax} 
becomes the probability distribution 
\begin{eqnarray}
 {\cal R}_N(r)  = N M_N \pi(r M_N ) \left[ 1- C(r M_N)\right]^{N-1}
\label{pdfratiomax}
\end{eqnarray}
Since the typical value $M_N$ is large as a consequence of the equation $C(M_N)=\frac{1}{N}$,
  while the ratio $r$ is finite,
 the value of $(r M_N)$ is also large,
i.e. the value of $C(r M_N ) $ is also small .
Then Eq. \ref{pdfratiomax} becomes 
\begin{eqnarray}
 {\cal R}_N(r) \opsimeq_{N \to +\infty} N M_N \pi(r M_N ) e^{ \displaystyle - N  C(r M_N ) }
\label{pdfrecmax}
\end{eqnarray}
It is thus more convenient to substitute $N=\frac{1}{C(M_N)}$
in order to write everything in terms of the single scale $ M_N$
\begin{eqnarray}
 {\cal R}_N(r)  \opsimeq_{N \to +\infty} 
 \frac{M_N  \pi(r M_N ) }{C( M_N )   } e^{ \displaystyle - \frac{  C(r M_N ) }{C( M_N )   } }
\label{wmax}
\end{eqnarray}

\subsection{ Case of the exponential decay }

For the exponential decay of Eq. \ref{expalpha},
the asymptotic behavior of the function $C(x)$ given in Eq. \ref{expcumul}
yields in Eq. \ref{wmax}
\begin{eqnarray}
 {\cal R}_N(r)  \opsimeq_{N \to +\infty} 
\alpha M_N^{\alpha} r^{\nu-1} e^{ M_N^{\alpha}(1-r^{\alpha}) } e^{ \displaystyle - r^{\nu-\alpha} e^{ M_N^{\alpha}(1-r^{\alpha}) } }
\label{wmaxexp}
\end{eqnarray}
This means that the rescaled variable
\begin{eqnarray}
y \equiv M_N^{\alpha} ( r^{\alpha} -1 ) + (\alpha-\nu) \ln r \opsimeq_{M_N \to +\infty}  M_N^{\alpha} ( r^{\alpha} -1 )
\label{y}
\end{eqnarray}
is distributed with the Gumbel probability distribution \cite{Gum,Gal}
\begin{eqnarray}
 G(y) \equiv e^{-y}  e^{ \displaystyle -  e^{-y} }
\label{gumbel}
\end{eqnarray}
whose very strong asymmetry for the asymptotic behaviors for $y \to \pm \infty$ is well-known
\begin{eqnarray}
 G(y) && \opsimeq_{y \to +\infty}  e^{-y}  
\nonumber \\
G(y) && \opsimeq_{y \to -\infty}    e^{ \displaystyle -  e^{-y} }
\label{gumbelasym}
\end{eqnarray}
As a consequence when $r$ is finite and different from the typical value $r^{typ}=1$,
the variable $y$ of Eq. \ref{y} will be near $(\pm \infty)$ depending on the sign of $(r-1)$ :
the asymptotic behaviors of Eqs \ref{gumbelasym}
will thus produce completely different scalings in the region bigger than typical $r >1$ \cite{maxmath,vivo}
\begin{eqnarray}
 {\cal R}_N(r)  \oppropto_{N \to +\infty}   e^{- M_N^{\alpha}  (  r^{\alpha} -1 ) } \ \ \ \ \ {\rm for } \ r>1
\label{pdfrbigger}
\end{eqnarray}
and in the region smaller than typical $0<r<1$
\begin{eqnarray}
 {\cal R}_N(r)  \oppropto_{N \to +\infty} 
 e^{ \displaystyle - r^{\nu-\alpha} e^{ M_N^{\alpha} ( 1-r^{\alpha} ) } } \ \ \ \ \ {\rm for } \ 0<r<1
\label{pdfrsmaller}
\end{eqnarray}

The link with the region of small typical fluctuations around the typical value $r^{typ}=1$
usually considered in the Extreme Value Statistics \cite{Gum,Gal}
corresponds here to the Taylor expansion at first order of the variable $y$ of Eq. \ref{y}
\begin{eqnarray}
y \opsimeq_{r \to 1} M_N^{\alpha} \left[ \alpha ( r -1 ) + O(r-1)^2) \right]
\label{ytaylor}
\end{eqnarray}
This yields that the appropriate rescaling 
to have a finite variable $y$ distributed with the Gumbel distribution $G(y)$ is
\begin{eqnarray}
r= 1+ \frac{y} {\alpha M_N^{\alpha}}
\label{ry}
\end{eqnarray}
or equivalently for the unrescaled maximum of Eq. \ref{empimax}
\begin{eqnarray}
x_N^{max} 
= r M_N =  \left( 1+ \frac{y} {\alpha M_N^{\alpha}} \right) M_N
=  M_N + \frac{y} {\alpha} M_N^{1-\alpha}
\label{recmax}
\end{eqnarray}
where the behavior of $M_N$ as a function of $N$ was recalled in Eq. \ref{mntypexp}.

\subsection{ Case of the power-law decay }

For the power-law decay of Eq. \ref{powermu},
the asymptotic behavior of its primitive of Eq. \ref{cumulpower}
yields in Eq. \ref{wmax}
\begin{eqnarray}
 {\cal R}_N(r)  \opsimeq_{N \to +\infty} 
  \frac{\mu }{ r^{1+\mu}  } e^{ \displaystyle - \frac{  1 }{ r^{\mu}   } } \equiv F_{\mu}(r)
\label{wmaxpower}
\end{eqnarray}
where the Fr\'echet distribution $F_{\mu}(r) $ 
of parameter $\mu$ appears for any finite value $r$,
while the scale $M_N$ has completely disappeared,
in contrast to the exponential decay case described above.
So here the probability of any value $r \ne r^{typ} \ne 1$ does not even decay with $N$.

\subsection{ Asymmetry beyond the regime of finite ratio $r$ }

In the following section, we will need the probability of the maximum of Eq \ref{pxmax} 
beyond the regime 
of finite ratio $r$ with respect to the typical value $M_N$.
In the region much bigger than typical 
$x^{max}_N \gg M_N $ where $C(x^{max}_N ) \ll C(M_N)=\frac{1}{N}$,
the factor $\left[ 1- C(x^{max}_N)\right]^{N-1} $ can be neglected in
Eq \ref{pxmax} and one obtains the leading behavior
\begin{eqnarray}
 {\cal X}_N(x^{max}_N)  \opsimeq_{ x^{max}_N \gg M_N }  N  \pi(x^{max}_N ) 
\label{xmaxbeyond}
\end{eqnarray}
The physical meaning is that one just needs to draw the anomalously big value $x^{max}_N $,
while the other $(N-1)$ variables may remain typical and thus have no probabilistic cost.

On the contrary, in the region much smaller
 than typical $x^{max}_N \ll M_N $ where $C(x^{max}_N ) \gg C(M_N)=\frac{1}{N}$,
the factor $\left[ 1- C(x^{max}_N)\right]^{N-1} $ is the leading behavior in
Eq \ref{pxmax} and produces an extensive cost in $N$ in the exponential
\begin{eqnarray}
 {\cal X}_N(x^{max}_N)  \opsimeq_{ x^{max}_N \ll M_N }  \left[ 1- C(x^{max}_N)\right]^{N-1}
\opsimeq_{ x^{max}_N \ll M_N }  e^{ - NC(x^{max}_N)  }
\label{xmaxsmall}
\end{eqnarray}

This asymmetry is thus very strong,
and reads for the exponential decay of Eq. \ref{expalpha} with Eq. \ref{expcumul},
\begin{eqnarray}
 {\cal X}_N(x^{max}_N)  && 
\opsimeq_{ x^{max}_N \gg M_N }   K N  ( x^{max}_N )^{\nu-1} e^{ -( x^{max}_N)^{\alpha}}
\nonumber \\
 {\cal X}_N(x^{max}_N)  && 
\opsimeq_{ x^{max}_N \ll M_N }  
e^{ - N  \frac{K}{\alpha} (x^{max}_N)^{\nu-\alpha} e^{- (x^{max}_N)^{\alpha} }  }
\label{xmaxasymexp}
\end{eqnarray}
while for the power-law decay of Eq \ref{powermu} with Eq \ref{cumulpower},
it is given by
\begin{eqnarray}
 {\cal X}_N(x^{max}_N)  && 
\opsimeq_{ x^{max}_N \gg M_N }   \frac{ N K}{ (x^{max}_N)^{1+\mu} }
\nonumber \\
 {\cal X}_N(x^{max}_N)  && 
\opsimeq_{ x^{max}_N \ll M_N }  e^{ - \frac{ N K}{(x^{max}_N)^{\mu}  }  }
\label{xmaxasympower}
\end{eqnarray}


\section{ Possible asymmetry in the large deviations of the empirical average  }

\label{sec_average}

Whenever the first moment $\overline{x} $ and
the variance $\sigma^2 \equiv \overline{x^2} - ( \overline{x} )^2$ are finite,
the Central Limit Theorem means that the empirical average of Eq \ref{empiav}
will display typical fluctuations
 of order $\frac{1 }{ \sqrt{N} } $ around the typical value $\overline{x}  $
\begin{eqnarray}
a_N \equiv \frac{1}{N} \sum_{i=1}^N x_i \opsimeq_{N \to +\infty} 
\overline{x} + \frac{v }{ \sqrt{\sigma^2 N} }
\label{clt}
\end{eqnarray}
where $v$ is a Gaussian random variable of zero mean and variance unity.
In this section, we focus on the large deviations properties
for the probability distribution ${\cal A}_N(a) $ of the empirical average $a$ 
\begin{eqnarray}
 {\cal A}_N(a) && \equiv
 \int_{0}^{+\infty} dx_1 \pi(x_1) ... \int_{0}^{+\infty} dx_N \pi(x_N) 
 \delta\left( a - \frac{1}{N} \sum_{i=1}^N x_i  \right)
\label{Anconvolution}
\end{eqnarray}
to discuss how rare it is to observe $a \ne  \overline{x}$ for large $N$ and when an asymmetry will occur.


\subsection{ Standard large deviation theory for the exponential decay with exponent $\alpha \geq 1$  }

The standard large deviation theory recalled in section \ref{sec_standardLargedev}
for additive empirical observable can be applied to the empirical average $G=a$ with $g(x)=x$ :
the probability to observe the value $a \ne a^{typ}=\overline{x}$ is exponentially small in $N$
\begin{eqnarray}
{\cal A}_N(a )  && \opsimeq_{N \to \infty} e^{ -N I (a)} 
\label{largedeav}
\end{eqnarray}
The rate function $I(a)$ can be either evaluated directly
or can be computed as the Legendre transform (Eqs \ref{legendre} and \ref{legendrereci}) of the
scaled cumulant generating function of Eq. \ref{phig}
\begin{eqnarray}
 \phi(k)  && = \ln \left(  \int_0^{+\infty} dx \pi(x)  e^{ k x }     \right)
\label{phikav}
\end{eqnarray}

For the exponential decay of Eq. \ref{expalpha} with an exponent $\alpha>1$,
the scaled cumulant generating function of Eq. \ref{phikav}
is defined for any $k \in ]-\infty,+\infty[$, and the above large deviation theory
can be applied without worry.

For the exponential decay of Eq. \ref{expalpha} with an exponent $\alpha=1$,
the scaled cumulant generating function of Eq. \ref{phikav}
is defined only for $k \in ]-\infty,1[$, and it is thus useful to
describe the example of the gamma distribution of parameter $\nu$ 
\begin{eqnarray}
\gamma_{\nu}(x) \equiv  \frac{1}{\Gamma(\nu)} x^{\nu-1} e^{- x} 
\label{gammanu}
\end{eqnarray}
of Laplace transform 
\begin{eqnarray}
\int_0^{+\infty} dx e^{-px} \gamma_{\nu}(x)  && = \frac{ 1 }{ (1+p)^{\nu} } 
\label{gammanulaplace}
\end{eqnarray}
So computing its power $N$ simply amounts to change the parameter $\nu$ into $(N \nu)$.
As a consequence, the sum of $N$ variables $x_i$ is distributed 
with the gamma distribution $\gamma_{N\nu} (.) $
of parameter $(N \nu)$ that corresponds to the convolution of $N$ distributions $ \gamma_{\nu}(.) $.
 After the rescaling by $N$, the probability distribution of the empirical average is
thus exactly given by
\begin{eqnarray}
{\cal A}_N(a )  && = N \gamma_{N\nu} (Na) = N \frac{1}{\Gamma(N\nu)} (Na)^{N\nu-1} e^{- Na} 
\label{gammaconvolution}
\end{eqnarray}
For large $N$, the Stirling approximation for $\Gamma(N\nu) $
yields the large deviation form of Eq. \ref{largedeav}
where the rate function 
\begin{eqnarray}
I_{\nu} (a) = a -\nu - \nu \ln \left( \frac{a}{\nu} \right) 
\label{rategamma}
\end{eqnarray}
 is well defined for $a \in ]0,+\infty[$
and measures how rare it is to observe a value $a$ different from the typical value $a^{typ}=\nu$.

The corresponding scaled generating cumulant function $\phi(k)$ of Eq. \ref{phikav}
is defined only for $k<1$.
\begin{eqnarray}
 \phi_{\nu}(k) = - \nu \ln (1-k )
\label{phiknu}
\end{eqnarray}
 The correspondence between $k$ and $a$ via the Legendre transform of Eq. \ref{legendre}
is
\begin{eqnarray}
 k_a =  I_{\nu}'(a) = 1 - \frac{\nu}{a}
\label{legendreka}
\end{eqnarray}
or equivalently via the reciprocal Legendre transform of Eq. \ref{legendrereci}
\begin{eqnarray}
 a_k   = \phi_{\nu}'(k)  = \frac{\nu} {1-k}
\label{legendrerecia}
\end{eqnarray}
So the region $k \in ]-\infty,0[$ allows to 
parametrize the whole smaller than typical region $ a \in ]0,\nu[$,
while the region $k \in ]0,1[$ allows to parametrize the whole bigger than typical region $ a \in ]\nu,+\infty[$ without problems.


\subsection{ Asymmetry in the large deviations for stretched exponential decay $0<\alpha<1$  }


\subsubsection{ Usual large deviation form in the region smaller than typical $a \leq a^{typ}=\overline{x} $  }

For the exponential decay of Eq. \ref{expalpha} with an exponent $0<\alpha<1$,
the scaled cumulant generating function of Eq. \ref{phikav}
is defined only for $k \in ]-\infty,0]$ that corresponds to
the region smaller than typical $a \leq a^{typ}=\overline{x}$,
where the usual large deviation form will thus be valid
\begin{eqnarray}
 {\cal A}_N(a ) \oppropto_{N \to +\infty}  e^{ - N I_-(a) } \ \ {\rm for } \ \ a \leq a^{typ}=\overline{x}
\label{ratesmaller}
\end{eqnarray}
The rate function $I_-(a)$ for $a \leq a^{typ}=\overline{x} $ corresponds to the Legendre transform 
of the function $\phi(k)$ defined for $k \leq 0$.


\subsubsection{ Unusual large deviation in the region $a>a^{typ}$    }

For $k > 0$ that corresponds to the region bigger than the typical value $a > a^{typ}=\overline{x}$,
the function $\phi(k)$ as defined by Eq. \ref{phikav} does not exist 
as a consequence of the divergence of the integral
at $(+\infty)$ when $\pi(x)$ decays only as a stretched exponential with $0<\alpha<1$
\begin{eqnarray}
 \int^{+\infty} dx \pi(x) e^{kx}  \propto  \int^{+\infty} dx  e^{kx- x^{\alpha}} =+\infty
\label{phikdv}
\end{eqnarray}

This suggests to consider the strategy based on the maximum alone :
one considers that $(N-1)$ variables have their typical sum $(N-1) \overline{x}$,
which happens with probability one for large $N$, i.e. with no probabilistic cost,
while the remaining variable, that will have to coincide with the maximum $x_N^{max}$
of Eq. \ref{empimax}, should be anomalously big
in order to satisfy the sum constraint
\begin{eqnarray}
x^{max}_N =  N a - (N-1) \overline{x} \oppropto_{N \to +\infty}  N( a- \overline{x})
\label{singlemax}
\end{eqnarray}
So the cost of this strategy directly involves the probability $ {\cal X}_N(x^{max}_N) $ of Eqs \ref{xmaxbeyond} and \ref{xmaxasymexp}
of the anomalously extensive value of Eq. \ref{singlemax}
\begin{eqnarray}
 {\cal A}^{StrategyMax}_N(a ) 
 \oppropto_{N \to +\infty} 
 {\cal X}_N(x^{max}_N \simeq N( a- \overline{x})
)  \opsimeq_{ x^{max}_N \gg M_N } 
\simeq   K N^{\nu}  ( a- \overline{x})^{\nu-1} e^{ - N^{\alpha} ( a- \overline{x})^{\alpha}}
\oppropto_{N \to +\infty} e^{ - N^{\alpha} I_+(a) } 
\label{single}
\end{eqnarray}
that decays only as the stretched exponential of exponent $\alpha \in ]0,1[$.
The corresponding rate function
\begin{eqnarray}
 I_+(a) = ( a- \overline{x})^{\alpha} \ \ {\rm for } \ \ a > \overline{x}
\label{iplusa}
\end{eqnarray}
has been proven to be valid in the whole region $a > \overline{x}$ in Refs \cite{nagaev,nina}.

\subsection{ Asymmetry in the large deviations for power-law decay   }

For the power-law decay of Eq. \ref{powermu}, one has the same scenario
as for the stretched exponential case discussed above:

(i) in the region smaller than typical $a \leq a^{typ}=\overline{x}$,
the usual large deviation form is valid
\begin{eqnarray}
 {\cal A}_N(a ) \oppropto_{N \to +\infty}  e^{ - N I_-(a) } \ \ {\rm for } \ \ a \leq a^{typ}=\overline{x}
\label{ratesmallerpower}
\end{eqnarray}
where the rate function $I_-(a)$ corresponds to the Legendre transform 
of the function $\phi(k)$ defined for $k \leq 0$.

(ii) in the region bigger than the typical value $a > a^{typ}=\overline{x}$
where the function $\phi(k)$ of Eq. \ref{phikav} does not exist 
as a consequence of the divergence of the integral,
the strategy based on the anomalous maximum of Eq. \ref{singlemax}
leads to the probability (using Eqs \ref{xmaxbeyond} and \ref{xmaxasympower})
\begin{eqnarray}
 {\cal A}^{StrategyMax}_N(a ) 
 \oppropto_{N \to +\infty} 
 {\cal X}_N(x^{max}_N \simeq N( a- \overline{x})
)  \opsimeq_{ x^{max}_N \gg M_N } 
\simeq   \frac{ K} { N^{\mu} ( a- \overline{x})^{1+\mu} }
\label{singlepower}
\end{eqnarray}
that decays only as the power-law $N^{-\mu}$.
This phenomenon of 'condensation' in the power-law case
has been studied in great detail in the references \cite{evans2008,godreche,evans2014},
with motivations coming from the zero-range process (see explanations and references in \cite{evans2008,godreche,evans2014}). Other physical applications can be found in 
\cite{barkai1,barkai2}.


\section{ Asymmetry in the large deviations of the empirical moment of order $q>0$  }

\label{sec_momentsq}

The analysis of the previous section concerning the empirical average 
can be directly generalized to obtain the large deviations properties
of the empirical moment of arbitrary non-integer order $q>0$ of Eq \ref{empimomentsq}.

\subsection{ Standard large deviation theory for the exponential decay with exponent $\alpha \geq q$  }

The standard large deviation theory recalled in section \ref{sec_standardLargedev}
for additive empirical observable can be applied to the empirical moment $G_N=a^{(q)}_N $
of arbitrary non-integer order $q>0$ of Eq \ref{empimomentsq} with $g(x)=x^q$ :
the probability to observe a value $a_q \ne \overline{x^q}$ is exponentially small in $N$
\begin{eqnarray}
{\cal A}^{(q)}_N(a_q )  && \opsimeq_{N \to \infty} e^{ -N I_q (a_q)} 
\label{largedeq}
\end{eqnarray}
and the rate function $I(a_q)$ corresponds to the Legendre transform (Eqs \ref{legendre} and \ref{legendrereci}) of the
scaled cumulant generating function of Eq. \ref{phig}
\begin{eqnarray}
 \phi_q(k)  && = \ln \left(  \int_0^{+\infty} dx \pi(x)  e^{ k x^q }     \right)
\label{phikq}
\end{eqnarray}
which is well defined for any $k \in ]-\infty,+\infty[$
when $\pi(x)$ displays the exponential decay of Eq. \ref{expalpha} with an exponent $\alpha>q$.

\subsection{ Asymmetry in the large deviations for stretched exponential decay $0<\alpha<q$  }


\subsubsection{ Usual large deviation form in the region smaller than typical $a_q \leq \overline{x^q} $  }

For the exponential decay of Eq. \ref{expalpha} with an exponent $0<\alpha<q$,
the scaled cumulant generating function of Eq. \ref{phikq}
is defined only for $k \in ]-\infty,0]$ that corresponds to
the region smaller than typical $a_q \leq \overline{x^q}$,
where the usual large deviation form will thus be valid
\begin{eqnarray}
 {\cal A}_N(a_q ) \oppropto_{N \to +\infty}  e^{ - N I_q^-(a) } \ \ {\rm for } \ \ a \leq \overline{x^q}
\label{ratesmallerq}
\end{eqnarray}
The rate function $I_q^-(a_q)$ corresponds to the Legendre transform 
of the function $\phi_q(k)$ defined for $k \leq 0$.


\subsubsection{ Unusual large deviation in the region $a_q> \overline{x^q}$    }

For $k > 0$ that corresponds to the region bigger than the typical value $a_q > \overline{x^q}$
the function $\phi_q(k)$ as defined by Eq. \ref{phikq} does not exist.
The strategy based on the maximum alone explained in the previous section 
can be then considered:
$(N-1)$ variables $x_i^q$ have their typical sum $(N-1) \overline{x^q}$,
which happens with probability one for large $N$, i.e. with no probabilistic cost,
while the remaining variable will have to coincide with 
the power $q$ of the maximum $x_N^{max}$
of Eq. \ref{empimax}.  This maximum should be anomalously big
in order to satisfy the sum constraint
\begin{eqnarray}
x^{max}_N = \left[  N a_q - (N-1) \overline{x^q} \right]^{\frac{1}{q} }
\oppropto_{N \to +\infty}  N^{\frac{1}{q} } ( a_q- \overline{x^q})^{\frac{1}{q} }
\label{singlemaxq}
\end{eqnarray}
So the cost of this strategy reads in terms of the probability $ {\cal X}_N(x^{max}_N) $ of Eqs \ref{xmaxbeyond} and \ref{xmaxasymexp}
of the anomalously big value of Eq. \ref{singlemaxq}
\begin{eqnarray}
 {\cal A}^{(q) StrategyMax}_N(a_q ) 
&& \oppropto_{N \to +\infty} 
 {\cal X}_N \left( x^{max}_N \simeq    
 N^{\frac{1}{q} } ( a_q- \overline{x^q})^{\frac{1}{q} } \right)
  \opsimeq_{ x^{max}_N \gg M_N } 
\simeq   K N^{1+ \frac{\nu-1}{q}}   ( a_q- \overline{x_q})^{\frac{\nu-1}{q}} 
e^{ - N^{\frac{\alpha}{q}} ( a_q- \overline{x^q})^{\frac{\alpha}{q}}}
\nonumber \\
&&
\oppropto_{N \to +\infty} e^{ - N^{\frac{\alpha}{q}} I_q^+(a_q) } 
\label{singleq}
\end{eqnarray}
that decays only as the stretched exponential of exponent $\frac{\alpha}{q} \in ]0,1[$,
with the corresponding rate function
\begin{eqnarray}
 I_q^+(a_q) = ( a_q- \overline{x^q})^{\frac{\alpha}{q}}  \ \ {\rm for } \ \ a_q > \overline{x^q}
\label{iplusaq}
\end{eqnarray}

\subsubsection{ Discussion    }

So for any exponential decay with exponent $\alpha>0$ in Eq. \ref{expalpha},
only the empirical moments of order $q<\alpha$ display a standard form of large deviations,
while the empirical moments of order $q>\alpha$ will be characterized by asymmetric large deviations.
For instance for the gamma distribution of Eq. \ref{gammanu} corresponding to $\alpha=1$,
where the large deviations of the empirical average $a$ corresponding to $q=1$ are still standard
(with the rate function of Eq. \ref{rategamma}), all the empirical moments of order $q>1$
will have asymmetric large deviations, in particular the empirical second moment 
corresponding to $q=2$.


\section{ Renormalization interpretation of large deviations rate functions }

\label{sec_rg}

The region of typical fluctuations around the typical value (see the Introduction around Eq. \ref{vtyp})
has been analyzed in detail from the renormalization point of view, 
both for the sum of random variables \cite{jona1,jona2,calvosum}
and for the maximum of random variables \cite{extreme1,extreme2,extreme3,extreme4,extreme5}.
In this section, it is thus interesting to discuss the meaning of large deviations
from the renormalization perspective.

\subsection{  Merging two sets of $N$ variables}

To see more clearly the renormalization meaning of large deviations,
it is interesting to consider the merging of two sets of $N$ random variables :

(1) the first set $[x_i]_{1 \leq i \leq N}$ of variables is drawn with the probability
distribution $\pi_1(x)$ and is characterized by the empirical histogram
\begin{eqnarray}
p^{(1)}_N(x) \equiv \frac{1}{N} \sum_{i=1}^N \delta(x-x_i)
\label{empihisto1}
\end{eqnarray}

(2) the second set $[x_i]_{N+1 \leq i \leq 2N}$ of variables is drawn with the probability
distribution $\pi_2(x)$ and is characterized by its empirical histogram
\begin{eqnarray}
p^{(2)}_N(x) \equiv \frac{1}{N} \sum_{i=N+1}^{2N} \delta(x-x_i)
\label{empihisto2}
\end{eqnarray}

\subsection{  Renormalization for the large deviations of the empirical histogram }

Each of these two sets labelled by $b=1,2$ is characterized by 
the large deviation properties of its empirical histogram $p^{(b)}_N(x)  $ (Eq. \ref{sanov} and \ref{s1relative})
\begin{eqnarray}
{\cal P}^{(b)}_N [ p_N^{(b)}(.)  ]   \opsimeq_{N \to +\infty} \delta \left( 1-  \int dx p^{(b)}_N(x)    \right)
e^{ \displaystyle - N S^{rel} ( p^{(b)}_N(.)  \vert \pi_b(.))  }
\label{sanov1}
\end{eqnarray}
or its exact generating function ${\cal Z}^{(b)}_N [ \kappa(.)  ] $ of Eq. \ref{geneq1} for any finite $N$
\begin{eqnarray}
 {\cal Z}^{(b)}_N [ \kappa(.)  ]  && 
\equiv  \int {\cal D}{ p^{(b)}_N(.) }  {\cal P}^{(b)}_N [ p^{(b)}_N(.)  ]    \ 
e^{ \displaystyle  N \int dx \kappa(x) p_N^{(b)}(x) }   
 = \left(  \int dx \pi_b(x)  e^{\kappa(x) }     \right)^N \equiv e^{ N \Phi_b [ \kappa(.) ]  }
\label{geneq11}
\end{eqnarray}

Via the merging of the data of Eqs \ref{empihisto1} and \ref{empihisto2},
the global histogram for the $(2N)$ variables is simply the average of the two histograms
\begin{eqnarray}
p_{2N}(x) \equiv \frac{1}{2N} \sum_{i=1}^{2N} \delta(x-x_i) = \frac{p^{(1)}_N(x) + p^{(2)}_N(x)}{2}
\label{empihistomerge}
\end{eqnarray}
with the typical value 
\begin{eqnarray}
p_{2N}^{typ}(x) = \frac{\pi_1(x) + \pi_2(x)}{2}
\label{empihistotypmerge}
\end{eqnarray}
Its generating function is simply the products of the generating functions of Eq. \ref{geneq11}
\begin{eqnarray}
 {\cal Z}_{2N} [ \kappa(.)  ] = {\cal Z}^{(1)}_N [ \kappa(.)  ] {\cal Z}^{(2)}_N [ \kappa(.)  ] 
= \left(  \int dx \pi_1(x)  e^{\kappa(x) }     \right)^N 
\left(  \int dx \pi_2(x)  e^{\kappa(x) }     \right)^N 
= e^{ N (\Phi_1 [ \kappa(.) ] +\Phi_2 [ \kappa(.) ]  )  } \equiv e^{2 N \Phi [ \kappa(.) ]  }
\label{geneq11merge}
\end{eqnarray}
i.e. the scaled cumulant generating function follows the renormalization rule
\begin{eqnarray}
\Phi [ \kappa(.) ] = \frac{ \Phi_1 [ \kappa(.) ] +\Phi_2 [ \kappa(.) ]  }{2}
\label{rgscaled}
\end{eqnarray}
In particular, when the two sets are drawn with the same probability distribution $\pi_1=\pi_2=\pi$,
the scaled cumulant generating function $\Phi [ \kappa(.) ] $ is exactly conserved along the RG flow.

In terms of large deviations form of Eq. \ref{sanov1},
the probability of the histogram of Eq. \ref{empihistomerge}
\begin{eqnarray}
 {\cal P}_{2N} [ p_{2N}(.)  ] && = \int {\cal D}{ p^{(1)}_N(.) }  {\cal P}^{(1)}_N [ p^{(1)}_N(.)  ]
 \int {\cal D}{ p^{(2)}_N(.) }  {\cal P}^{(2)}_N [ p^{(2)}_N(.)  ]
\delta \left( p_{2N}(.) - \frac{p^{(1)}_N(.) + p^{(2)}_N(.)}{2} \right)
\nonumber \\
&&   \opsimeq_{N \to +\infty} \int {\cal D}{ p^{(1)}_N(.) } \int {\cal D}{ p^{(2)}_N(.) }
\delta \left( 1-  \int dx p^{(1)}_N(x)    \right)\delta \left( 1-  \int dx p^{(2)}_N(x)    \right)
\delta \left( p_{2N}(.) - \frac{p^{(1)}_N(.) + p^{(2)}_N(.)}{2} \right)
\nonumber \\
&& \ \ \ \ \ e^{ \displaystyle  N \left[ - S^{rel} ( p^{(1)}_N(.)  \vert \pi_1(.)) 
- S^{rel} ( p^{(2)}_N(.)  \vert \pi_2(.)) \right]
 }
\label{sanovmerge}
\end{eqnarray}
corresponds to the optimization of the function $\left[ - S^{rel} ( p^{(1)}_N(.)  \vert \pi_1(.)) 
- S^{rel} ( p^{(2)}_N(.)  \vert \pi_2(.)) \right] $ in the exponential
in the presence of the constraints that can be taken into account via Lagrange multipliers.
One obtains the optimal solution
\begin{eqnarray}
p^{(1)}_N(x)  && = p_{2N}(x) \frac{ 2 \pi_1(x) }{ \pi_1(x)+\pi_2(x) }
\nonumber \\
p^{(2)}_N(x)  && = p_{2N}(x) \frac{ 2 \pi_2(x) }{ \pi_1(x)+\pi_2(x) }
\label{sanovmergeopt}
\end{eqnarray}
and the corresponding sum of the relative entropies in the exponential
\begin{eqnarray}
&&   N \left[ - S^{rel} ( p^{(1)}_N(.)  \vert \pi_1(.)) - S^{rel} ( p^{(2)}_N(.)  \vert \pi_2(.)) \right]
  = N \left[ - \int dx  p_N^{(1)}(x)  \ln \left( \frac{ p_N^{(1)}(x)  }{ \pi_1(x)   } \right)
 - \int dx  p_N^{(2)}(x)  \ln \left( \frac{ p_N^{(2)}(x)  }{ \pi_2(x)   } \right)  \right]
\nonumber \\
&& = - 2N \int dx p_{2N}(x) \ln \left( \frac{ 2 p_{2N} (x)  }{  \pi_1(x)+\pi_2(x)  } \right)
= - 2N S^{rel} \left( p_{2N} (x) \big \vert p_{2N}^{typ}(x) =\frac{\pi_1(x)+\pi_2(x)}{2}  \right)
\label{sanovmergerelative}
\end{eqnarray}
coincides with the relative entropy of the histogram $p_{2N} (x) $ with respect to its typical value of Eq. \ref{empihistotypmerge}
as it should for consistency.
In particular, when the two sets are drawn with the same probability distribution $\pi_1=\pi_2=\pi$,
the optimal solution to produce an anomalous empirical histogram $p_{2N}(x) $
consists in choosing the same anomalous empirical histogram  for the two subsets (Eq. \ref{sanovmergeopt}).

\subsection{ Renormalization for the large deviations of the empirical maximum  }

For each set $b=1,2$, the cumulative probability distribution of the empirical maximum
of Eq. \ref{pxmax}
can be interpreted as an exact large deviation form
\begin{eqnarray}
  \int_0^x dx' {\cal X}_{N}^{(b)}(x') =   \left[1- C_b(x) \right]^N = e^{- N J_b(x) }
\label{cumulmaxrgb}
\end{eqnarray}
where the rate function reads
\begin{eqnarray}
J_b(x) = -  \ln \left[1- C_b(x) \right]
\label{ratejb}
\end{eqnarray}
in terms of the complementary cumulative distribution function (Eq \ref{cumulative})
associated to each distribution $\pi_b(x)$
\begin{eqnarray}
C_b(x) \equiv \int_{x }^{+\infty}  dx' \pi_b(x')
\label{cumulativeb}
\end{eqnarray}

The empirical maximum of the $(2N)$ variables is of course the maximum
of the two maximal values associated to the two sets of Eqs \ref{empihisto1} and \ref{empihisto2}
\begin{eqnarray}
x^{max}_{2N} \equiv \max   \left( x^{max(1)}_N , x^{max(2)}_N  \right)
\label{empimaxrg}
\end{eqnarray}
So the corresponding cumulative distribution 
\begin{eqnarray}
 \int_0^{x} dx' {\cal X}_{2N}(x')  =  \left[ \int_0^x dx' {\cal X}_{N}^{(1)}(x') \right]
\left[ \int_0^x dx' {\cal X}_{N}^{(2)}(x') \right] = e^{- N \left[ J_1(x)+J_2(x) \right] } \equiv e^{- 2N J(x)}
\label{cumulmaxrg}
\end{eqnarray}
is written exactly in a large deviation form with the rate function
\begin{eqnarray}
 J(x) = \frac{ J_1(x)+ J_2(x) }{2} = -  \frac{ \ln \left[1- C_1(x) \right] + \ln \left[1- C_2(x) \right] }{2} 
\label{ratej}
\end{eqnarray}
When the two sets are drawn with the same probability distribution $\pi_1=\pi_2=\pi$,
one obtains that the rate function $J(x)$ 
 is exactly conserved along the RG flow.

\subsection{Renormalization for the large deviations of the empirical average  }

\subsubsection{Case of standard large deviations   }

In the case of standard large deviations, each set $b=1,2$ is 
described by the large deviation form of Eq. \ref{largedeav}) for its empirical average $a_b$
\begin{eqnarray}
{\cal A}^{(b)}_N(a_b )  && \opsimeq_{N \to \infty} e^{ -N I_b (a_b)} 
\label{largedeavb}
\end{eqnarray}
where the rate function $I_b(a_b)$ is the Legendre transform
of the
scaled cumulant generating function of Eq. \ref{phikav}
\begin{eqnarray}
 \phi_b(k)  && = \ln \left(  \int_0^{+\infty} dx \pi_b(x)  e^{ k x }     \right)
\label{phikavb}
\end{eqnarray}
involved in the generating function
\begin{eqnarray}
 Z_{N}^{(b)} (k)  && 
\equiv   \int da_b {\cal A}^{(b)}_{N}(a_b) 
e^{ \displaystyle   N k a_b }    \equiv e^{ N \phi_b(k)  }
\label{zbav}
\end{eqnarray}

Via the merging of the data of Eqs \ref{empihisto1} and \ref{empihisto1},
the empirical average of the $(2N)$ variables is simply the average of the two empirical averages
of the two sets $b=1,2$
\begin{eqnarray}
a \equiv \frac{1}{2N} \sum_{i=1}^{2N} x_i= \frac{ a_1+a_2}{2}
\label{rgempiav}
\end{eqnarray}
Its generating function is simply the products of the generating functions of Eq. \ref{zbav}
\begin{eqnarray}
 Z_{2N} (k)  && 
\equiv  \int da {\cal A}_{2N}(a) e^{   2 N k a }   
= \int da_1 {\cal A}^{(1)}_{N}(a_1) \int da_2 {\cal A}^{(2)}_{N}(a_2)   \ 
e^{   2 N k (a_1+a_2) }  
\nonumber \\
&&  = Z_N^{(1)}(k) Z_N^{(2)}(k) 
= e^{ N \left[ \phi_1(k)+\phi_2(k) \right] } \equiv e^{ 2N \phi(k)  }
\label{genegk}
\end{eqnarray}
so the renormalization rule for the 
 scaled cumulant generating function is simply
\begin{eqnarray}
\phi(k) = \frac{ \phi_1(k)+\phi_2(k)   }{2}
\label{rgscaledk}
\end{eqnarray}
In particular, when the two sets are drawn with the same probability distribution $\pi_1=\pi_2=\pi$,
the scaled cumulant generating function $\phi(k)  $ is exactly conserved along the RG flow.

In terms of large deviations form of Eq. \ref{largedeavb},
the probability of the empirical average reads for this case $\pi_1=\pi_2=\pi $ 
\begin{eqnarray}
 {\cal A}_{2N} (a) && = \int da_1 {\cal A}_N(a_1 ) \int da_2 {\cal A}_N(a_2 ) 
\delta  \left( a - \frac{ a_1+a_2}{2} \right)
 \opsimeq_{N \to + \infty} 
\int da_1   e^{ -N \left[ I (a) - I(2a-a_1) \right] } 
\label{distriavmerge}
\end{eqnarray}
The saddle-point evaluation of this integral requires to find the maximum of the function in the exponential
\begin{eqnarray}
{\cal L} (a_1) = -  I (a_1) - I(2a-a_1) 
\label{lavmerge}
\end{eqnarray}
The vanishing of the first derivative
\begin{eqnarray}
0 = \frac{ \partial {\cal L} (a_1)}{\partial a_1}  = -  I' (a_1) + I'(2a-a_1) 
\label{lavmergederi}
\end{eqnarray}
gives the symmetric solution $a_1=a=2a-a_1=a_2$ 
which is indeed a maximum if the second derivative is negative 
\begin{eqnarray}
0 = \frac{ \partial^2 {\cal L} (a_1)}{\partial^2 a_1} \vert_{a_1=a} = \left[ -  I''(a_1) - I''(2a-a_1) \right]_{a_1=a} = - 2 I'' (a) <0
\label{lavmergederi2}
\end{eqnarray}
For instance for the gamma distribution of parameter $\nu$ of Eq. \ref{gammanu}
the second derivative of the rate function of Eq. \ref{rategamma} 
\begin{eqnarray}
I_{\nu}'' (a) = \frac{ \nu }{a^2} >0
\label{rategammaderi2}
\end{eqnarray}
satisfies this condition.

\subsubsection{Case of large deviations with asymmetric scaling  }

It is now interesting to compare the above discussion
with the case of the large deviations for stretched exponential decay $0<\alpha<1$  
that display the asymmetric scaling (Eqs \ref{ratesmaller} and \ref{single})
\begin{eqnarray}
 {\cal A}_N(a ) && \oppropto_{N \to +\infty}  e^{ - N I_-(a) } \ \ {\rm for } \ \ a \leq \overline{x}
\nonumber \\
 {\cal A}_N(a ) && \oppropto_{N \to +\infty} e^{ - N^{\alpha} I_+(a) } \ \ {\rm for } \ \ a > \overline{x}
\label{asymb}
\end{eqnarray}
Then Eq. \ref{distriavmerge}
is replaced by the sum of four possible contributions of various orders with respect to $N$
\begin{eqnarray}
 {\cal A}_{2N} (a) && = \int da_1
\left[ e^{ - N I_-(a_1) } \theta( a_1 \leq \overline{x}) 
+e^{ - N^{\alpha} I_+(a_1) } \theta( \overline{x} < a_1 ) \right]
 \nonumber \\ && 
\left[ e^{ - N I_-(2a-a_1) } \theta( 2a-\overline{x} \leq a_1)
 +e^{ - N^{\alpha} I_+(2a-a_1) } \theta( a_1 <2a-\overline{x}  ) \right] 
\nonumber \\ && 
= \int da_1
 e^{ - N \left[ I_-(a_1) + I_-(2a-a_1)\right] } 
\theta( 2a-\overline{x} \leq a_1 \leq \overline{x})
\nonumber \\ && 
+ \int da_1
 e^{ - N I_-(a_1)- N^{\alpha} I_+(2a-a_1)  } \theta( a_1 \leq \overline{x})  \theta( a_1 <2a-\overline{x}  ) 
\nonumber \\ && 
+ \int da_1
e^{ - N^{\alpha} I_+(a_1) - N I_-(2a-a_1) } \theta( \overline{x} < a_1 )  \theta( 2a-\overline{x} \leq a_1)
\nonumber \\ && 
+\int da_1
e^{ - N^{\alpha} \left[ I_+(a_1)+ I_+(2a-a_1) \right]  } \theta( \overline{x} < a_1 <2a-\overline{x}  ) 
\label{distriavmergetheta}
\end{eqnarray}
The fourth contribution of order $e^{- N^{\alpha} }$ will be the leading contribution
whenever the domain of integration for $a_1$ is not empty, i.e. in the region $a> \overline{x}$ :
then the saddle-point evaluation requires the maximization of the function involving 
the rate function $I_+(a)=(a-\overline{x}  )^{\alpha}$ of Eq. \ref{iplusa}
\begin{eqnarray}
{\cal L} (a_1) = -  I_+ (a_1) - I_+(2a-a_1)
\label{lavmergeplus}
\end{eqnarray}
However the symmetric solution $a_1=a=2a-a_1=a_2$ is a minimum here as a consequence of the sign of the second derivative for any $0<\alpha<1$
\begin{eqnarray}
{\cal L}'' (a) = - 2 I_+'' (a) = 2 \alpha (1-\alpha) (a-\overline{x}  )^{\alpha-2} >0
\label{lavmergeplusderi}
\end{eqnarray}
The maximization of Eq \ref{lavmerge} occurs instead at the boundaries 
$a_1= \overline{x}$ and $a_2=2a-  \overline{x}$ (or vice-versa)
and one obtains the leading contribution in the region $a>\overline{x} $
\begin{eqnarray}
 {\cal A}_{2N} (a) && \opsimeq
e^{ - N^{\alpha}  I_+(2a-\overline{x} )   } 
= e^{ - N^{\alpha}  (2a-2 \overline{x} )^{\alpha}  } 
= e^{ - (2N)^{\alpha}  (a- \overline{x} )^{\alpha}   } 
= e^{ - (2N)^{\alpha} I_+(a)  } \ \ {\rm for } \ \ a > \overline{x}
\label{distriavmergethetabig}
\end{eqnarray}
as it should for consistency with Eq. \ref{asymb}  in the region $a>\overline{x} $.


\section{ Conclusion }

\label{sec_conclusion}

In this paper,  we have revisited the empirical observables based on independent random variables, namely the empirical maximum, the empirical average, the empirical non-integer moments or other additive empirical observables, in order to describe the cases where asymmetric large deviations occur. We have stressed the analogy with equilibrium statistical mechanics : the Sanov theorem for the large deviations of the empirical histogram that involves as rate function the relative entropy with respect to the true probability distribution
has been taken as the unifying starting point. The various empirical observables have been then analyzed by optimizing this relative entropy in the presence of the appropriate constraints. 

Finally, we have discussed the physical meaning of large deviations rate functions from the renormalization perspective. While most renormalization procedures
have been studied in the past at the level of their typical fluctuations, it will be thus interesting in the future to re-analyze them at the level of their large deviations,
 as in the recent study \cite{c_largedevrg} concerning disordered directed polymers
 where asymmetric large deviations are known to occur,

\end{document}